%% file: main.tex
\documentclass[conference]{IEEEtran}
\IEEEoverridecommandlockouts
\usepackage{cite}
\usepackage{amsmath,amssymb,amsfonts}
\usepackage{algorithmic}
\usepackage{graphicx,import}
\usepackage{textcomp}
\usepackage{xcolor}
\usepackage{xspace}
\usepackage{adjustbox}
\usepackage{scalefnt}
\usepackage{pgf}
\usepackage{lmodern}
\usepackage{listings}
\usepackage{booktabs}
\usepackage{subcaption}
\usepackage{paralist}
\usepackage[linesnumbered,lined,boxed,commentsnumbered]{algorithm2e}
\PassOptionsToPackage{hyphens}{url}\usepackage[hidelinks]{hyperref}

\def\BibTeX{{\rm B\kern-.05em{\sc i\kern-.025em b}\kern-.08em
    T\kern-.1667em\lower.7ex\hbox{E}\kern-.125emX}}
    
\definecolor{backcolour}{rgb}{0.95,0.95,0.92}
\lstdefinestyle{mystyle}{
    backgroundcolor=\color{backcolour}, 
    basicstyle=\ttfamily\scriptsize,
    breakatwhitespace=false,         
    breaklines=true,                 
    captionpos=b,                    
    keepspaces=true,                 
    showspaces=false,                
    showstringspaces=false,
    showtabs=false,                  
    tabsize=2
}
\newcommand{\wse}{WSE\xspace}
\newcommand{\wsetwo}{WSE-2\xspace}

\newcommand{\fd}{Finite Differences\xspace}

\begin{document}
\title{Massively scalable stencil algorithm}

\author{
\IEEEauthorblockN{Mathias Jacquelin\textsuperscript{\textsection}}
\IEEEauthorblockA{\textit{Cerebras Systems Inc.} \\
Sunnyvale, California, USA\\
{mathias.jacquelin@cerebras.net}}
\and
\IEEEauthorblockN{Mauricio Araya-Polo\textsuperscript{\textsection} and Jie Meng}
\IEEEauthorblockA{
\textit{TotalEnergies EP Research \& Technology US, LLC.}\\
Houston, Texas, USA\\
mauricio.araya@totalenergies.com}}

\maketitle
\begingroup\renewcommand\thefootnote{\textsection}
\footnotetext{Equal contribution.}
\endgroup
\begin{abstract}

Stencil computations lie at the heart of many scientific and industrial applications. 
Unfortunately, stencil algorithms perform poorly on machines with cache based memory hierarchy, due to low re-use of memory accesses. This work shows that for stencil computation a novel algorithm that leverages a localized communication strategy effectively exploits the Cerebras WSE-2, which has no cache hierarchy. This study focuses on a 25-point stencil finite-difference method for the 3D wave equation, a kernel frequently used in earth modeling as numerical simulation. In essence, the algorithm trades memory accesses for data communication and takes advantage of the fast communication fabric provided by the architecture. 
The algorithm ---historically memory bound--- becomes compute bound.
This allows the implementation to achieve near perfect weak scaling, reaching up to 503 TFLOPs on \wsetwo, a figure that only full clusters can eventually yield.

\end{abstract}


\section{Introduction}
Stencil computations are central to many scientific problems and industrial applications, from weather forecast~(\cite{Thaler2019}) to earthquake modeling~(\cite{moczo2014}). The memory access pattern of this kind of algorithm,
 in which all values in memory are accessed but used in only very few arithmetic operations, 
is particularly unfriendly to hierarchical memory systems of traditional architectures. 
Optimizing these memory operations is the main focus of performance improvement research on the topic.

Subsurface characterization is another area where stencils are widely used. The objective is to identify major structures in the subsurface that can either hold hydrocarbon or be used for $CO_2$ sequestration. One step towards that end is called seismic modeling, where artificial perturbations of the subsurface are modeled solving the wave equation for given initial and boundary conditions. Solving seismic modeling efficiently is crucial for subsurface characterization, since many perturbation sources need to be modeled as the subsurface model iteratively improves. The numerical simulations required by seismic algorithms for field data are extremely demanding, falling naturally in the HPC category and requiring  practical evaluation of technologies and advanced hardware architectures to speed up computations. 

Advances in hardware architectures have motivated algorithmic changes and optimizations to stencil applications for at least 20 
years~(\cite{Rivera2000}).
Unfortunately, the hierarchical memory systems of most current architectures is not well-suited
to stencil applications, therefore limiting performance. This applies to multi-core machines, clusters of multi-cores, and accelerator-based platforms such as GPGPUs, FPGAs, etc.~(\cite{datta2009,araya2011}).
Alternatively, non-hierarchical architectures were explored in this context, such as the IBM Cell BE~(\cite{araya2009}), yielding high computational efficiency but with limited impact.

A key element for large scale simulations is the potential of deploying substantial number of processing units connected by an efficient 
fabric. The Cell BE lacked the former and it had limited connectivity. Another example of non-hierarchical memory system is 
the Connection Machine (\cite{kahle1989}), which excelled on scaling but at the cost of a very complex connectivity. In this work, a novel stencil algorithm based on localized communications that does not depend on memory hierarchy optimizations is introduced. This algorithm can take advantage of architectures such as the \wse from Cerebras (\cite{hotchips2019}) and potentially Anton 3-like systems (\cite{Anton3}). These are examples of architectures addressing both limitations described above.

\begin{table}
    \centering
    \begin{adjustbox}{width=.8\linewidth}
    \begin{tabular}{c|c}
        \hline
        \textbf{Traditional architecture} & \textbf{\wse}\\
        \hline
        L1 & Memory \\
        L2 \& L3 & $\varnothing$ \\
        DRAM & $\varnothing$\\
        Off-node interconnect & Fabric \& routers \\
        \hline
    \end{tabular}
    \end{adjustbox}
    \caption{Equivalences between traditional architectures and the \wse}
    \label{tab:wse_equiv}
    \vspace{-0.2cm}
\end{table}

Another angle to be considered is the availability of hardware-based solutions in the market. Literature review yields no generally available hardware architecture addressing the specific bottlenecks of stencil applications. Only a few custom designs examples are available~(\cite{Krueger11,STX19}).

In this work, an implementation of such seismic modeling method on a novel architecture is presented. The proposed mapping requires a complete redesign of the basic stencil algorithm. The contribution of this work is multi-fold:
\begin{itemize}
    \item An efficient distributed implementation of \fd for seismic modeling on a fully-distributed memory architecture with 850,000 processing elements.
    \item A stencil algorithm that is performance bound to the capacity of individual processing element rather than bound by memory or communication bandwidth.
    \item The target application ported relies on an industry-validated stencil order. 
\end{itemize}

The paper is organized as follows: Section~\ref{sec:related} reviews relevant contributions in the literature. Section~\ref{sec:target} describes the target application.
Section~\ref{sec:wse} provides details of how the target application was redesigned to efficiently use the novel processor architecture.
Sections~\ref{sec:eval} and \ref{sec:roof} discusses experimental results and profiling data. Section \ref{sec:conclu} provides discussions and conclusions.

\import{}{related_work.tex}

\import{}{fd.tex}

\import{}{fd_on_wse.tex}

\import{}{experiments.tex}

\import{}{roofline.tex}
\section{Conclusion}
\label{sec:conclu}

In this work, a \fd algorithm taking advantage of low-latency localized communications
and flat memory architecture has been introduced. 
Localized broadcast patterns are introduced
to exchange data between processing elements and fully utilize the interconnect.
Experiments show that it is possible to reach near perfect weak scalability on distributed memory architecture such as the \wsetwo. On this platform, the implementation of \fd reaches 503 TFLOPs. This is a remarkable throughput for this stencil order on a single node machine. The roofline model introduced in this work confirms that \fd becomes compute-bound on the \wsetwo. This demonstrates the validity and potential of the approach presented in this work, and demonstrate how different hardware architectures like the \wsetwo can be exploited efficiently by stencil-based applications.

Future efforts include the integration of the ported kernel at the center of more ambitions applications, such as the ones regularly used by seismic modeling experts when taking real-life decisions. Further, given the well established capacity of this hardware architecture for ML-based applications, a hybrid HPC-ML approach will also be investigated.

One interesting consequence of having a relatively compact machine delivering such a high performance level for this type of application is that seismic data processing can happen at the same time it is acquired on the field, which is key when constant monitoring is required. Furthermore, under this scenario, processing capacity can move from data centers closer to where sensors are, namely target edge-HPC. 
\section*{Acknowledgements}

Authors would like to thank Cerebras Systems and TotalEnergies
for allowing to share the material.
Also, authors would like to acknowledge Grace Ho, Ashay Rane, and Natalia Vassilieva from Cerebras for the contributions, and Ruychi Sai from Rice U. for fruitful discussions about GPGPU optimized kernels.  

\newpage

\bibliographystyle{IEEEtranS}
\bibliography{refs}
\end{document}

%% file: related_work.tex
\section{Related work}
\label{sec:related}
\subsection{Stencil Computation}

Not all stencil computations are the same, and the structure and order of the stencil set the limits of the attainable performance. The higher the order (neighbors to be accounted for) and the closer to a pure star shape is, the harder to compute the stencil is. Traditional hierarchical memory subsystems will be overwhelmed by the memory access pattern which displays very little data reuse.  
Considerable amount of research effort has been devoted to optimizing stencil computations, and to finding ways around these issues.
Spurred by emerging hardware technologies, studies on how stencil algorithms can be tailored to fully exploit unique hardware characteristics have covered many aspects, from  DSLs, performance modeling, to pure algorithmic optimizations, targeting low-level architectural features in some cases.

Domain-specific languages (DSLs), domain-specific parallel programming models, and compiler optimizations for stencils have been proposed (e.g., \cite{devito-api,gysi2020domainspecific,rawat2019optimizing, ghosh2012experiences}). Performance models have been developed for this computing pattern (see~\cite{DELACRUZ20112146,stengel15}), and the kernel has been ported to a variety of platforms (\cite{datta2009,araya2011,araya2009,Zhang2020}) including specific techniques to benefit from unique hardware features. 

Stencil computations have also been the subject of multiple algorithmic optimizations. Spatial and temporal blocking has been proposed \cite{wonnacott2000,frigo2005,Kronawitter2018}. A further example is the semi-stencil algorithm proposed by De la Cruz et al. \cite{delacruz2014semi}, which offers an improved memory access pattern and a higher level of data reuse. Promising results are also achieved using a higher dimension cache optimization, as introduced by Nguyen et al.~\cite{Nguyen2010} , accommodating both thread-level and data-level parallelism. Most recently, Sai et al. \cite{ryuichi_cpe_2021} studied high-order stencils with a manually crafted collection of implementations of a 25-point seismic modeling stencil in CUDA and HIP for the latest GPGPU hardware. Along this line of hardware-oriented stencil optimizations, Matsumura et al.~\cite{matsumura2020} proposed a framework (\emph{AN5D}) for GPU stencil optimization, obtaining remarkable results.

Wafer-scale computations have first been explored in Rocki et al.~\cite{rocki2020fast}, in which the authors explore a BiCGStab implementation to solve a linear system arising from a 7-point finite difference stencil on the first generation of Cerebras Wafer-Scale Engine. 
Albeit computing a much simpler stencil and having a higher arithmetic intensity, this study paved the way to the work presented in this study.
 A notable difference between the current work and this study is that floating point operations were performed in mixed precision: stencil and AXPY operations being computed using 16 bit floating point operations and global reductions using 32 bit arithmetic. In the present study, only 32 bit floating point arithmetic is used, and neither AXPY operation nor global reductions are involved. This makes the performance achieved by these two applications not directly comparable.

%% file: fd.tex
\section{Finite difference for seismic modeling}
\label{sec:target}
\textit{Minimod} is a proxy application that simulates the propagation of waves through the Earth models, by solving a Finite Difference (FD) which is discretized form of the wave equation. It is designed and developed by TotalEnergies EP Research \& Technologies~\cite{meng_minimod_2020}. Minimod is self-contained and designed to be portable across multiple compilers. The application suite provides both non-optimized and optimized versions of computational kernels for targeted platforms. The main purpose is benchmarking of emerging new hardware and programming technologies. Non-optimized versions are provided to allow analysis of pure compiler-based optimizations. 

In this work, one of the kernels contained in Minimod is used as target for redesign: the acoustic isotropic kernel in a constant-density domain~\cite{qawasmeh2017performance}. For this kernel, the wave equation PDE has the following form:

\begin{equation}
\frac{1}{\mathbf{V}^2}\frac{\partial^2 \mathbf{u}}{\partial t^2} - \nabla^2 \mathbf{u} = \mathbf{f}, 
\end{equation}

where $\mathbf{u} = \mathbf{u}(x,y,z)$ is the wavefield, $\mathbf{V}$ is the Earth model (with velocity as the main property), and $\mathbf{f}$ is the source perturbation. The equation is discretized in time using a $2^{nd}$ order centered stencil, resulting in the semi-discritized equation:

\begin{equation}
\mathbf{u}^{n+1} - \mathbf{Q}\mathbf{u}^n + \mathbf{u}^{n-1} = \left(\Delta t^2\right) \mathbf{V}^2 \mathbf{f}^n,\label{eq:minimod-semidisc}
\end{equation}
\begin{equation*}
	\mathrm{with\ }\mathbf{Q} = 2 + \Delta t^2 \mathbf{V}^2 \nabla^2.
\end{equation*}

Finally, the equation is discretized in space using a 25-point stencil in 3D ($8^{th}$ order in space), with four points in each direction as well as the centre point:
\begin{align*}
\nabla^2 \mathbf{u}(x,y,z) \approx \sum_{m=0}^4 &c_{xm}\left[\mathbf{u}(i+m,j,k) + \mathbf{u}(i-m,j,k)\right] &+ \\
									 &c_{ym}\left[\mathbf{u}(i,j+m,k) + \mathbf{u}(i,j-m,k)\right] &+ \\
									 &c_{zm}\left[\mathbf{u}(i,j,k+m) + \mathbf{u}(i,j,k-m)\right]
\end{align*}
where $c_{xm},c_{ym},c_{zm}$ are discretization parameters, solved in step 4 in Algorithm~\ref{algo:minimod}. In the remainder of the document we refer to this operator as the Laplacian.\\

A simulation in Minimod consists of solving the wave equation at each timestep for thousands of timesteps. Pseudocode of the algorithm is shown in Algorithm~\ref{algo:minimod}. 
\begin{algorithm}
	\KwData{$\mathbf{f}$: source}
	\KwResult{$\mathbf{u}^n$: wavefield at timestep $n$, for $n\leftarrow 1$ \KwTo $T$}
	$\mathbf{u}^0 := 0$\;
	\For{$n\leftarrow 1$ \KwTo $T$}{\nllabel{line:tsloop}
		\For{each point in wavefield $\mathbf{u}^n$}{
			Solve Eq.~\ref{eq:minimod-semidisc} (left hand side) for wavefield $\mathbf{u}^n$\;
		}
		$\mathbf{u}^n = \mathbf{u}^n + \mathbf{f}^n$ (Eq.~\ref{eq:minimod-semidisc} right hand side)\;
	}\nllabel{line:end-of-ts}
	\caption{Minimod high-level description}
	\label{algo:minimod}
\end{algorithm}

We note that  a full simulation includes additional kernels, such as I/O and boundary conditions. These additional kernels are not evaluated in this study but will be added in the future. The kernel has been ported and optimized for GPGPUs, including NVIDIA A100, full report can be found in \cite{sai2020accelerating}, this implementation is used as baseline to compare the results with the proposed implementation.

%% file: fd_on_wse.tex
\section{Finite-differences on the \wse}
\label{sec:wse}

This section introduces general architectural details of the Cerebras Wafer-Scale Engine (\wse) and discusses hardware features allowing the target application to reach the highest level of performance.
The mapping of the target algorithm onto the system is then discussed. The implementation is referred to as \fd in the remainder of the study. Communication strategy and core computational parts involved in \fd are also reviewed. 

The implementation of \fd on the \wse is written in Cerebras Software Language (\textit{CSL}) using the Cerebras SDK~\cite{sdk_arxiv}, which allows software developers to write custom programs for Cerebras systems. 
CSL is a C-like language based on Zig~\cite{Zig18website}, a reinterpretation of C which provides a simpler syntax and allows to declare compile-time blocks/optimizations explicitly (rather than relying on macros and the C preprocessor). 
CSL provides direct access to key hardware features, while allowing the use of higher-level constructs such as functions and while loops. The language allows to express computations and communications across multiple cores. Excerpts provided in the following will use the CSL syntax.

\subsection{The \wse architecture}

\begin{figure*}[tb]
\centering
\begin{adjustbox}{width=.8\linewidth}
\includegraphics{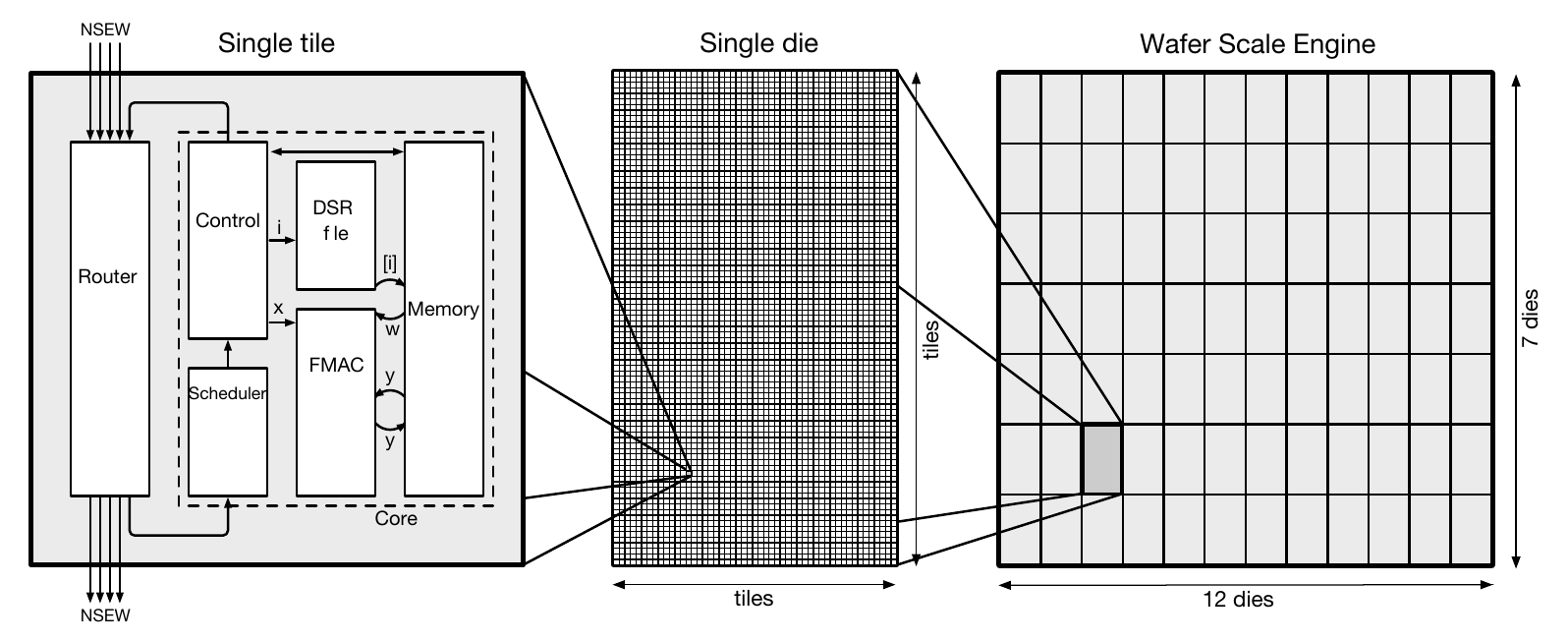}
\end{adjustbox}
\caption{
An overview of the Wafer Scale Engine (WSE). The \wse (to the right) occupies an entire wafer, and is a 2D array of dies. Each die is itself a grid of tiles (in the middle), which contains a router, a processing element and single-cycle access memory (to the left). In total, the \wse-2 embeds 2.6 trillion transistors in a silicon area of 46,225 $mm^2$ .
\label{fig.wse_archi}}
\end{figure*}

The \wse is an unprecedented-scale manycore processor. It is the first wafer-scale system~\cite{hotchips2019}, embedding all compute and memory resources within a single silicon wafer, together with a high performance communication interconnect. An overview of the architecture is given in Figure~\ref{fig.wse_archi}.
In its latest version, the \wsetwo provides a total of 850,000 processing elements, each with 48 KB of dedicated SRAM memory; up to eight 16-bit floating point operations per cycle; 16 bytes of read and 8 bytes of write bandwidth to the memory per cycle; and a 2D mesh interconnection fabric that can handle 4 bytes of bandwidth per PE per cycle in steady state~\cite{lie2021multi}.

The \wse can be seen as an on-wafer distributed-memory machine with a 2D-mesh interconnection \textit{fabric}. This on-wafer network connects \textit{processing elements} or PEs.
Each PE has a very fast local memory and is connected to a router. The routers link to the routers of the four neighboring PEs. There is no shared memory.
The \wse contains a $7 \times 12$ array of identical ``dies'', each holding thousands of PEs. Other chips are made by cutting the wafer into individual die. In the \wse, however, the interconnect is extended between dies. 
This results in a wafer-scale processor tens of times larger than the largest processors on the market at the time of its release.

The instruction set of the \wse is designed to operate on vectors or higher dimensionality 
objects. This is done by using \textit{data structure descriptors},
which contain information regarding how a particular object should be 
accessed and operated on (such as address, length, stride, etc.).

As mentioned above, given the distributed memory nature of the \wse, the interconnect plays a crucial role in delivering performance. It is convenient to think of the 2D mesh interconnect in terms of cardinal directions. Each PE has 5 full-duplex links managed by its local router: East, West, North, and South links allow data to reach other routers and PEs, while the \textit{ramp} link allows data to flow between the router and the PE, on which computations can take place. Each link is able to move a 32 bit packet in each direction per cycle. Each unidirectional link operates in an independent fashion, allowing concurrent flow of data in multiple directions.

Every 32 bit packet has a \textit{color} (contained in additional bits of metadata). The role of colors is twofold: 
\begin{asparaenum}
\item Colors are used in the routing of communications. A color determines the packet's progress at each router it encounters from source to destination(s). 
A router controls, for each color, where --- to what subset of the five links --- to send a packet of that color.
Moreover, colors are akin to virtual channels in which ordering is guaranteed.
\item Colors can also be used to indicate the type of a message: a color can be associated to a handler triggered when a packet of that particular color arrives. 
\end{asparaenum}





\begin{figure}
\centering
\begin{adjustbox}{width=.4\linewidth}
\scalefont{2}
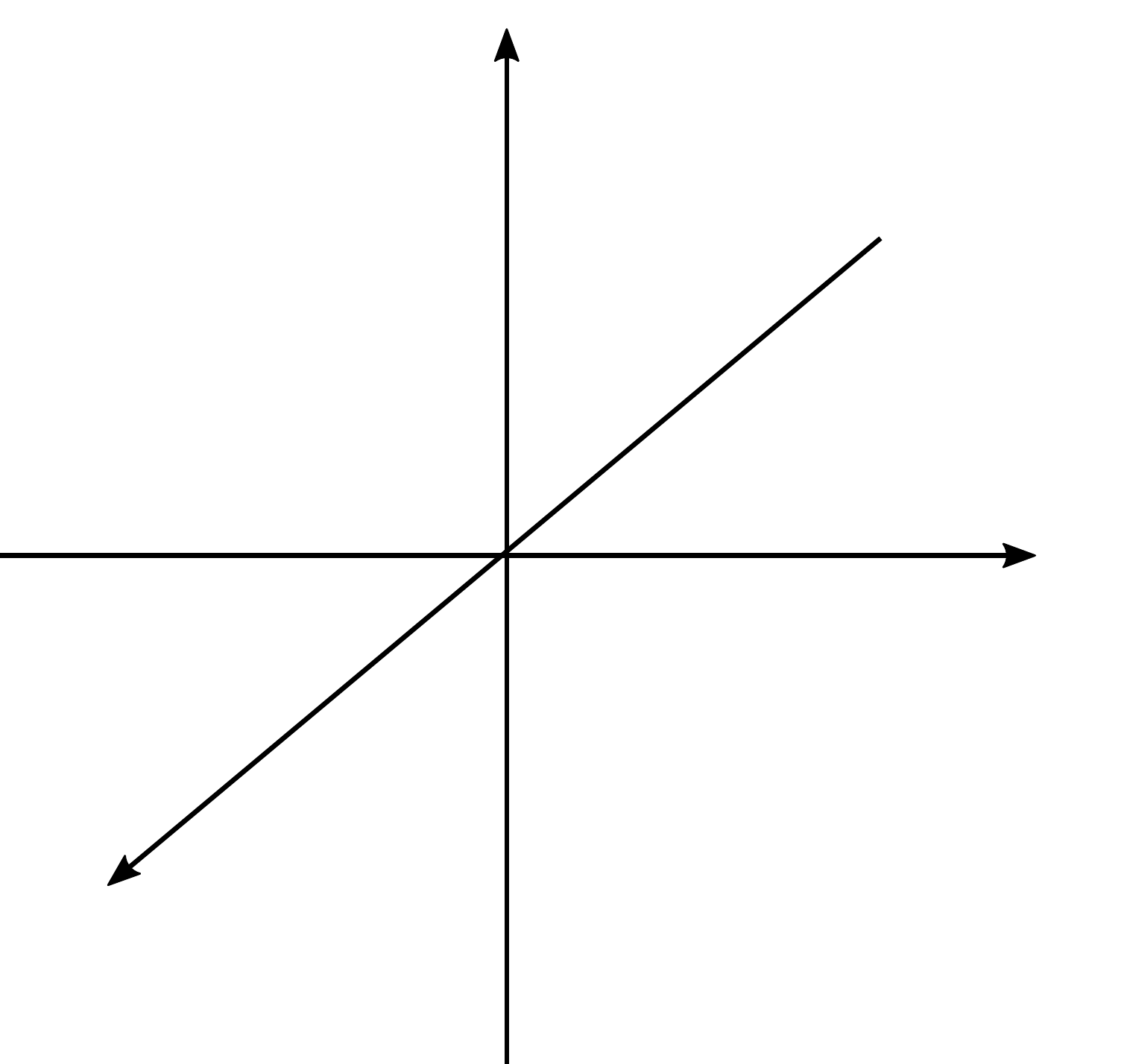
\end{adjustbox}
\caption{The 25-point stencil used in \fd. Cells in white (along $Z$) reside in the local memory of a PE, blue cells are communicated along the $X$ dimension, and green cells are communicated along the $Y$ dimension.}
\label{fig.stencil}
\end{figure}

The \wse is an unconventional parallel computing machine in the sense that the entire distributed memory machine lies within the same wafer. There is no cache hierarchy nor shared memory. Equivalences between hardware features of the \wse and what they correspond to on traditional architectures (such as distributed memory supercomputers) are summarized in Table~\ref{tab:wse_equiv}.



\subsection{Target Algorithm/application mapping}

\begin{figure}
\centering


\begin{subfigure}[t]{0.6\linewidth}
\centering
 \raisebox{-.05\height}{%
\begin{adjustbox}{width=\linewidth}%
\scalefont{3}%
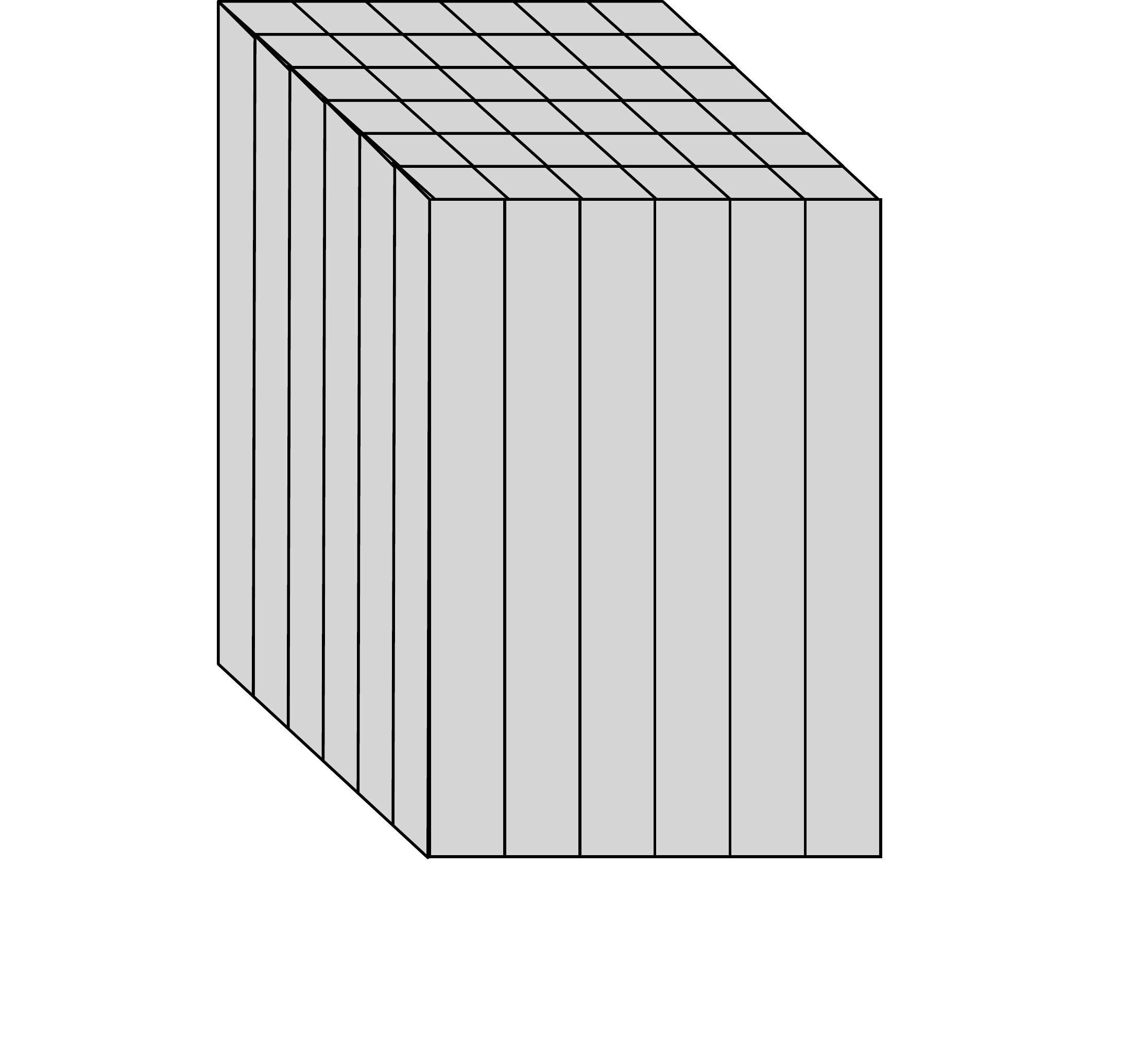%
\end{adjustbox}%
}
\caption{3D grid of size $nx \times ny \times nz$. $X$ and $Y$ dimensions are mapped onto the PE grid of the \wse.\label{fig.domain_mapping}}
\end{subfigure}
\hfill
\begin{subfigure}[t]{0.35\linewidth}
\centering
\begin{adjustbox}{width=.7\linewidth}
\scalefont{3}
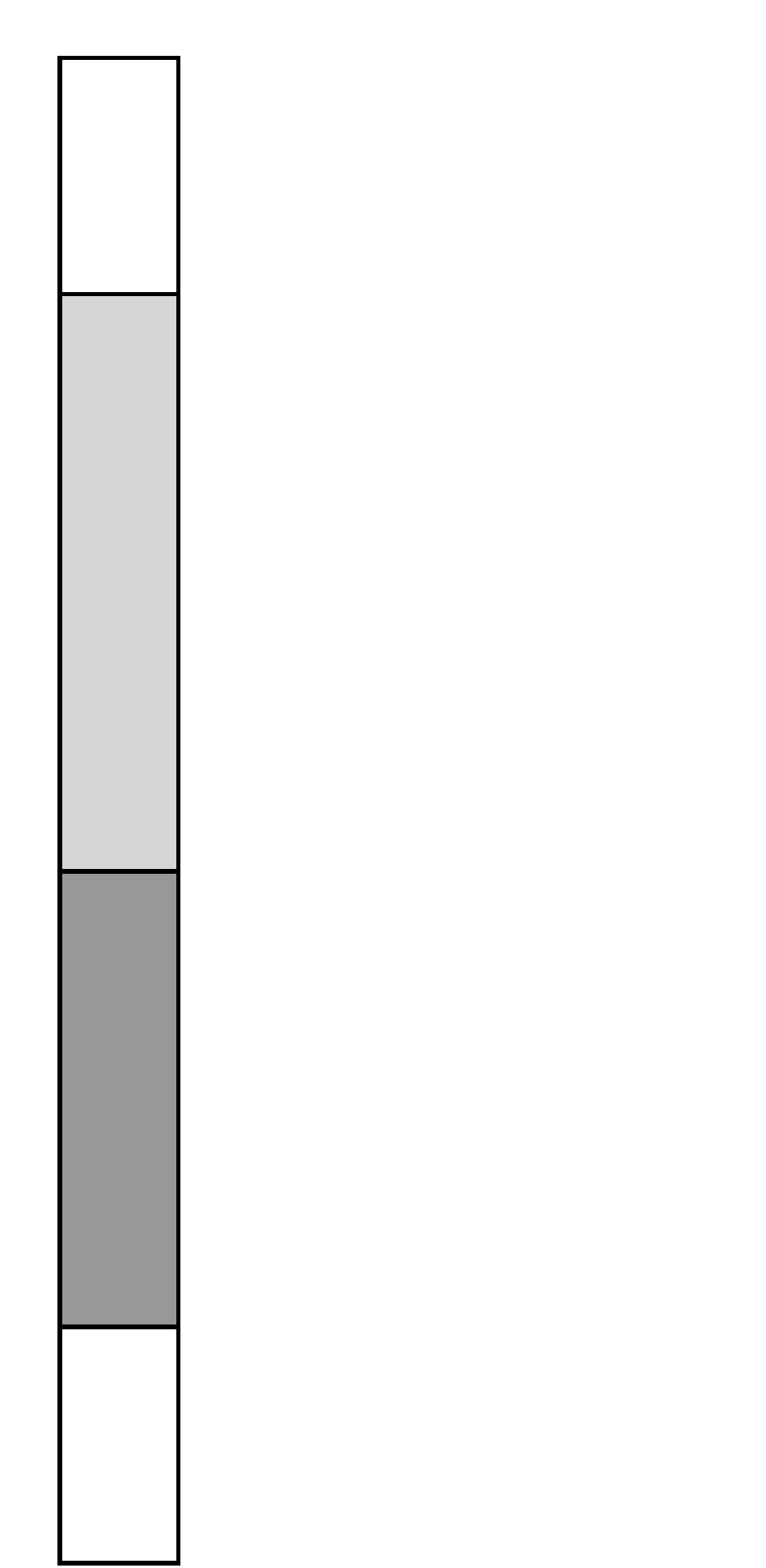
\end{adjustbox}
\caption{Column of cells stored in local memory, extended by 8 cells. Operations are carried out by blocks of size $b$.\label{fig.local_domain}}
\end{subfigure}
\caption{Computing pattern mapping}
\vspace{-0.2cm}
\end{figure}

The sheer scale of the \wse calls for a novel mapping of the target algorithm onto the processor. The 3D $nx \times ny \times nz$ grid on which the stencil computation is performed is mapped onto the \wse in two different ways: the $X$ and $Y$ dimensions are mapped onto the fabric while the $Z$ dimension is mapped into the local memory of a PE. This follows the approach that was explored in Rocki et al.~\cite{rocki2020fast}, and has the benefit of expressing the highest possible level of concurrency for this particular application. Figure~\ref{fig.domain_mapping} depicts how the domain is distributed over the \wse.
Each PE owns a subset of $nz$ cells of the original grid, as depicted in Figure~\ref{fig.local_domain}. In order to simplify the implementation, this local subset is extended by 8 extra cells: 4 cells below and 4 cells above the actual grid. This ensures that any cell in the original grid always has 4 neighbors below and 4 neighbors above. A PE stores the wavefield at two time steps (see Equation~\ref{eq:minimod-semidisc}). In order to lower overheads, computations and communications are performed on blocks of size $b$. The block size is chosen to be the largest such that the $2\times(nz + 8)$ cells and all buffers depending on $b$ can fit in memory.


\subsection{Stencil computation}
\label{subsec.stencil}
Computing the spatial component (referred to as the Laplacian) of the governing PDE lies at the heart of the target application, and it is traditionally the most demanding component. In addition to requiring a significant amount of floating point operations, computing the Laplacian also involves data movement, which is known to be very expensive on distributed memory platforms.

In the context of this paper, a 25-point stencil (depicted in Figure~\ref{fig.stencil}) is used. The stencil spans over all three dimensions of the grid. In order to compute a particular cell, data from neighboring cells is needed in all three dimensions. More precisely, a $\mathrm{\textit{cell}}_{x,y,z}$ requires data from neighboring cells: 
\begin{equation*}
\begin{split}
\mathrm{\textit{cell}}_{x-4\leq i<x,y,z}, ~~~~ \mathrm{\textit{cell}}_{x<i\leq x+4,y,z},\\
\mathrm{\textit{cell}}_{i,y-4\leq j<y,z}, ~~~~ \mathrm{\textit{cell}}_{i,y<j\leq y+4,z},\\
\mathrm{\textit{cell}}_{i, j, z-4\leq k<z}, ~~~~ \mathrm{\textit{cell}}_{i,j,z<k\leq z+4}.
\end{split}
\end{equation*}

\subsubsection{Localized broadcast patterns}

Dimensions $X$ and $Y$ from the grid are mapped onto the fabric of the \wse. To compute the stencil, a PE has therefore to communicate with 4 of its neighbors along each cardinal direction of the PE grid. 
A communication strategy similar to Rocki et al.~\cite{rocki2020fast}, in which a single color is used per neighboring PE, would have resulted in an excessive color use for the stencil of interest to this application.
In this work, \textit{localized broadcast patterns} along every PE grid directions (\textit{Eastbound} and \textit{Westbound} for the $X$ dimension, and \textit{Northbound} and 
\textit{Southbound} for the $Y$ dimension) are used instead. Each broadcast pattern uses two dedicated colors (one for receiving data, one for sending data) and can happen concurrently with others broadcast patterns using separate links to communicate between PEs. Given the stencil size used in the application and the number of colors available on the hardware, the limited color usage per broadcast pattern is critical to the feasibility of the implementation.

\begin{figure*}[tb]
\centering
\hfill
\begin{minipage}{0.3\linewidth}
\vspace{-4cm}
\centering
\begin{subfigure}{\linewidth}
\centering
\begin{adjustbox}{width=.6\linewidth}%
\scalefont{1}%
\includegraphics{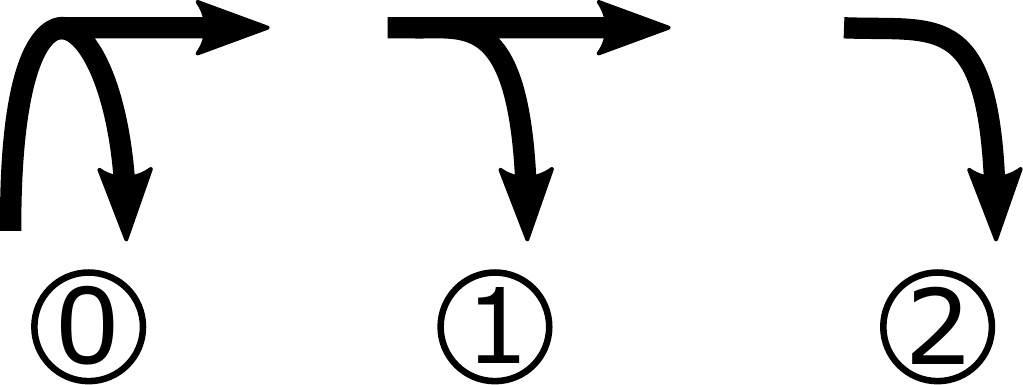}%
\end{adjustbox}%
\caption{
\wsetwo router configurations used by \fd. Configuration 0 corresponds to the configuration of the Root of a broadcast, configuration 1 is used by PEs in the middle, configuration 2 is used by the Last PE.\label{fig.switch_positions}}
\end{subfigure}
\end{minipage}
\hfill
\begin{subfigure}[t]{0.65\linewidth}
\centering
\begin{adjustbox}{width=.75\linewidth}
\scalefont{1}
\includegraphics{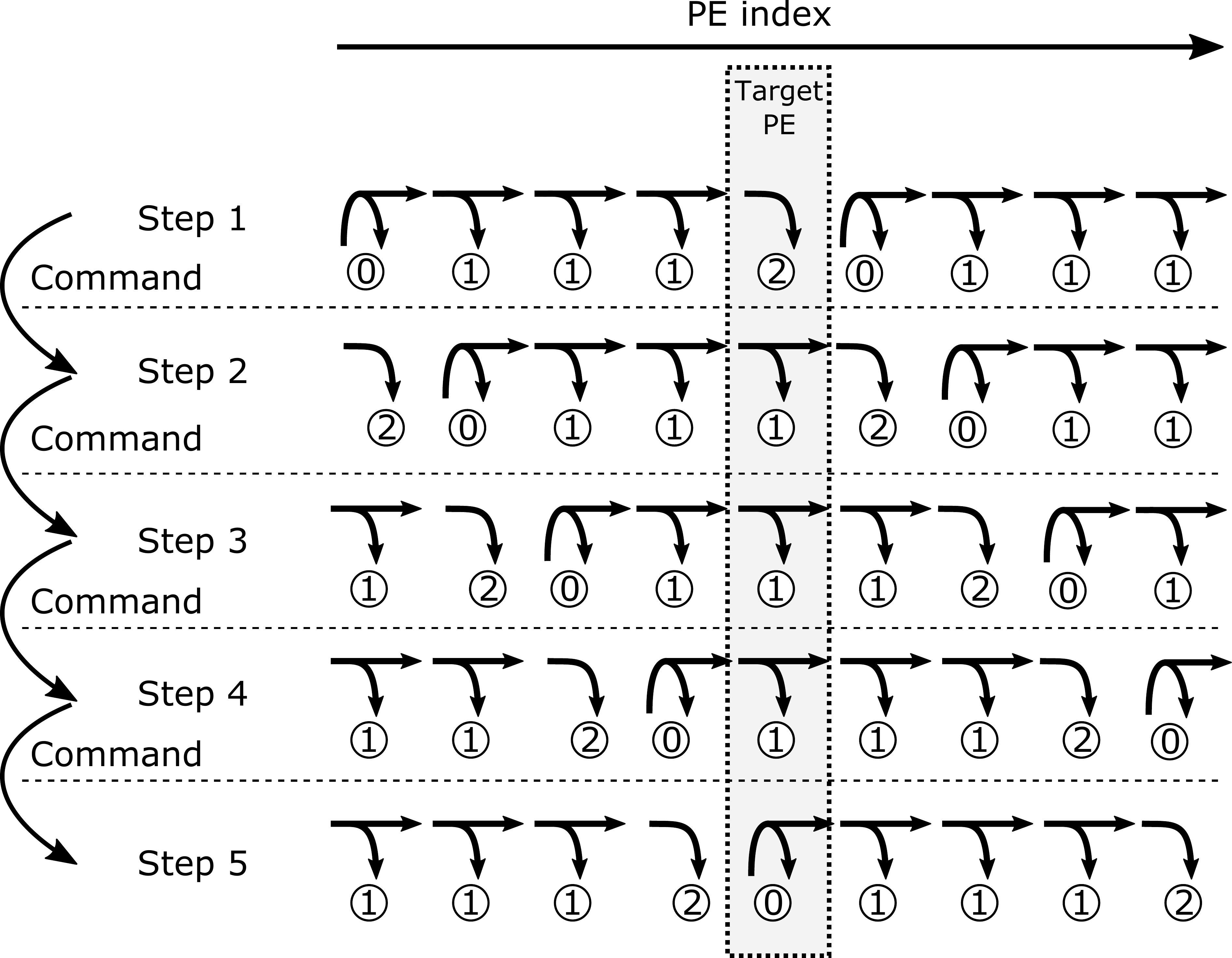}
\end{adjustbox}
\caption{5 communication steps required to fetch all the data required by a target PE from the West (steps 1 through 4) and to send its data to the East (step 5). Corresponding router configurations are given in the circled numbers. At each step, a router command is sent through the broadcast pattern, changing the configurations of each set of 5 routers.
\label{fig.repeated_broadcasts}}
\end{subfigure}
\hfill
\caption{Eastward \textit{localized broadcast} operation used in \fd to exchange cells along the $X$ dimension.}
\end{figure*}

In each broadcast pattern, multiple \textit{Root} PEs send their local block of data of length $b$ to their respective neighboring 4 processing elements. This pattern is depicted in Figure~\ref{fig.repeated_broadcasts} for the Eastward direction.

The router of each PE is configured to control how packets are received and transmitted. Each router determines, for each color, the incoming links from which that color can be received and the subset of the five outgoing links to which that color will be sent. The routing can be changed at run-time by special commands which can be sent just as other packets are sent.
This capability lies at the heart of the communication strategy proposed here.
In Figure~\ref{fig.switch_positions}, the different router configurations used by \fd are given. All Root PEs are in configuration 0. Intermediate PEs in each broadcast are in configuration 1, while Last PEs are in configuration 2.
Ideally, a Root PE should broadcast its data only to other PEs. However, due to hardware constraints, a Root PE is obliged to receive its own data as well.

After sending its local data, a Root PE sends a command to its local router and the following 4 routers. In effect, this routing update shifts the communication pattern by one position: the first neighbor now becomes a Root in the next step of the broadcast pattern.
After 5 steps (and 5 shifts), a PE has sent its data out and has received data from its 4 neighbors. In Figure~\ref{fig.repeated_broadcasts}, the \textit{target PE} receives data from the West during the first 4 steps, and sends its data to the East at step 5.

One of the very important aspects of this is that changing the routing on a remote router does not require any action from the local PE. It is therefore uninterrupted and can perform computations simultaneously. Another advantage is that all the control logic is encapsulated in this routing update. A particular PE has to do two things only: sending $b$ cells out, and receiving $5 \times b$ cells (from 4 neighbors and itself). The router configuration will determine when the data flows in or out of a given router. Once a PE is notified that its data has been sent out, it sends a router command to update the routing and transition to the next step of the broadcast pattern. There is no bookkeeping required to determine whether a PE is in a given position in a broadcast.

\subsubsection{Stencil computation over the $X$ and $Y$ dimensions}
In order to compute the stencil over the $X$ and $Y$ axes, communications between PEs are 
required. As the stencil involved in this application is a 25-point stencil, data from 16 
neighboring PEs along the $X$ and $Y$ directions must be exchanged. This means that at each time step, a PE is involved in 4 localized broadcast patterns (one per cardinal direction). In each broadcast pattern, a PE sends its data and receives data from 4 neighbors.

\begin{figure}
\centering
\begin{adjustbox}{width=.95\linewidth}
\scalefont{3}
\includegraphics{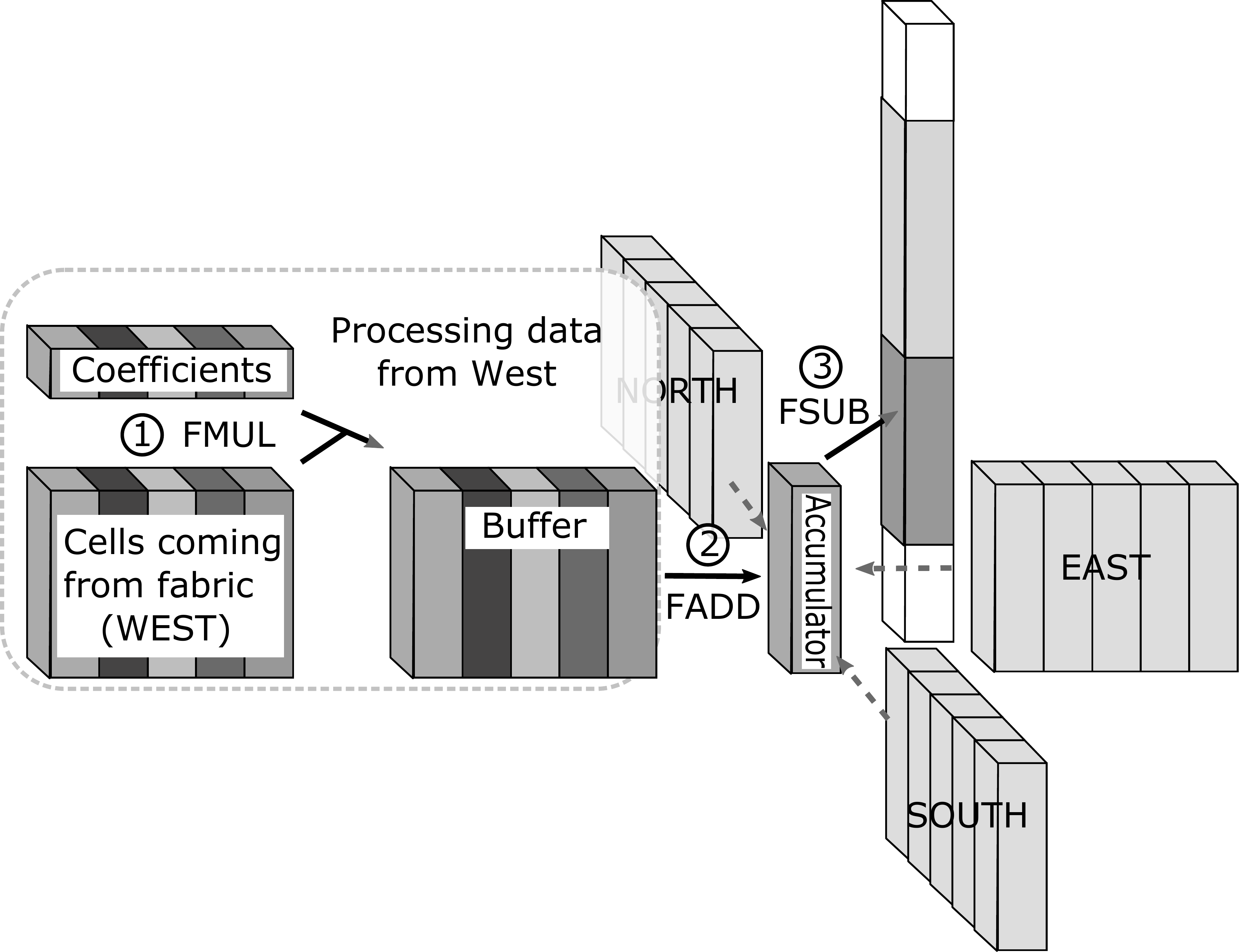}
\end{adjustbox}
\caption{A summary of main operations: computing the stencil over the $X$ and $Y$ dimensions (for each cardinal direction), reducing the \textit{accumulator} buffer, and subtracting the accumulator from the wavefield.
\label{fig.xy_update}}
\vspace{-0.2cm}
\end{figure}


Using a FMUL instruction, incoming cells from a given direction are multiplied ``on the fly'' with coefficients depending on their respective distance to the local cell. There are 4 FMUL operations happening concurrently (one per cardinal direction). This is depicted as step 1 in Figure~\ref{fig.xy_update} for the data coming from the West. Each FMUL instruction operates on $5 \times b$ incoming cells coming from a particular cardinal direction, and the coefficients (corresponding to $c_{xm}$ and $c_{ym}, ~ \forall m \in \{1 \ldots 4\}$ in Section~\ref{sec:target}). A given coefficient is applied to $b$ consecutive cells are they are coming from the same distance neighbor.

Since a PE is receiving data from itself, it is advantageous to compute the contribution from the center $\mathrm{\textit{cell}}_{x,y,z}$ during this step. This is done during the FMUL operation that processes the cells coming from the West, by multiplying the cells coming from the same PE with $c_{x0} + c_{y0} + c_{z0}$. FMULs operating on cells coming from all other directions use a coefficient of 0 for the data coming from the same PE.

Once this distributed computation phase is complete, the data of size $4 \times 5 \times b$ is reduced
into a single buffer of $b$ cells (which is referred to as \textit{accumulator}) using a FADD instruction (step 2 in Figure~\ref{fig.xy_update}). The dimension of size 4 corresponds to the number of localized broadcast patterns a PE participates in, 5 corresponds to the number of PE it is receiving from, and $b$ is the number of elements coming from each PE.
All contributions of neighboring cells along the $X$ and $Y$ dimensions are contained in the \textit{accumulator} buffer after the reduction.

\subsubsection{Stencil computation over the $Z$ dimension}

After remote contributions to the Laplacian from the $X$ and $Y$ axes of the grid have been accumulated, contributions from $Z$ can be computed.
Given the problem mapping over the \wse, this means that, at each time step, the computation 
over the $Z$ dimension can be performed in an embarrassingly parallel fashion since this dimension resides entirely in the memory of a PE. 

\begin{figure}
\centering
\begin{adjustbox}{width=.8\linewidth}
\scalefont{4}
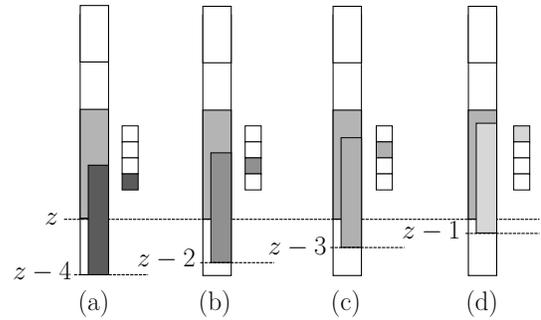
\end{adjustbox}
\caption{Applying the stencil over the $Z$ dimension
\label{fig.z_update}}
\vspace{-0.15cm}
\end{figure}

Each PE executes 8 FMACs instructions of length $b$, multiplying the wavefield by one of the 8
coefficients (corresponding to discretization parameters $c_{zm}, ~ \forall m \in {1 \ldots 4}$). The result of each FMAC is placed into the \textit{accumulator} buffer (which also 
contains the contributions from the $X$ and $Y$ dimensions).
Given a target block of size $b$ starting at coordinate $z_b$, each FMAC takes an input block starting at index $z_b + \mathrm{\textit{offset}}$ and multiplies it by a coefficient. 
The $\mathrm{\textit{offset}}$ values are $\{0, 1, 2, 3\}$ and $\{5, 6, 7 ,8\}$, and corresponding coefficients are $\{c_{z4}, c_{z3}, c_{z2}, c_{z1}\}$ and $\{c_{z1}, c_{z2}, c_{z3}, c_{z4}\}$. The CSL code is provided in Figure~\ref{lst.z_update} and the first 4 steps of this process are illustrated in Figure~\ref{fig.z_update}. As can be seen, this step skips offset 4, which would correspond to the multiplication by $c_{z0}$, since that particular computation has already been done as discussed earlier. At the end of this step, the Laplacian is contained in the \textit{accumulator} buffer.

\begin{figure*}
\centering
\begin{subfigure}{.4\linewidth}
\centering
\[
\begin{split}
    \mathrm{\textit{accumulator}}_{z_{b}\leq i<z_{b}+b} = ~\mathrm{\textit{accumulator}}_{z_{b}\leq i<z_{b}+b} \\
    + \mathrm{\textit{zWF}}_{z_{b}-4 \leq i < z_{b}+b-4} \times c_{z4}\\
    + \mathrm{\textit{zWF}}_{z_{b}-3 \leq i < z_{b}+b-3} \times c_{z3}\\
    + \mathrm{\textit{zWF}}_{z_{b}-2 \leq i < z_{b}+b-2} \times c_{z2}\\
    + \mathrm{\textit{zWF}}_{z_{b}-1 \leq i < z_{b}+b-1} \times c_{z1}\\
    + \mathrm{\textit{zWF}}_{z_{b}+1 \leq i < z_{b}+b+1} \times c_{z1}\\
    + \mathrm{\textit{zWF}}_{z_{b}+2 \leq i < z_{b}+b+2} \times c_{z2}\\
    + \mathrm{\textit{zWF}}_{z_{b}+3 \leq i < z_{b}+b+3} \times c_{z3}\\
    + \mathrm{\textit{zWF}}_{z_{b}+4 \leq i < z_{b}+b+4} \times c_{z4}
\end{split}
\]
	\caption{Operations performed along the $Z$ dimension. $\mathrm{\textit{zWF}}$ is the wavefield stored in the local memory of a PE}
\end{subfigure}
\hfill
\begin{subfigure}{.52\linewidth}
\begin{lstlisting}
const accumDsd = @get_dsd(mem1d_dsd, .{
  .tensor_access = |i|{nz} -> accumulator[i]});
const srcZ0 = @get_dsd(mem1d_dsd, .{
  .tensor_access = |i|{nz} -> zWF[0, i]});
@fmacs(accumDsd, accumDsd, srcZ0, coefficients[0]);
const srcZ1 = @increment_dsd_offset(srcZ0, 1, f32);
@fmacs(accumDsd, accumDsd, srcZ1, coefficients[1]);
const srcZ2 = @increment_dsd_offset(srcZ0, 2, f32);
@fmacs(accumDsd, accumDsd, srcZ2, coefficients[2]);
const srcZ3 = @increment_dsd_offset(srcZ0, 3, f32);
@fmacs(accumDsd, accumDsd, srcZ3, coefficients[3]);
// srcZ4 not used: update is done in Eastwards broadcast
const srcZ5 = @increment_dsd_offset(srcZ0, 5, f32);
@fmacs(accumDsd, accumDsd, srcZ5, coefficients[5]);
const srcZ6 = @increment_dsd_offset(srcZ0, 6, f32);
@fmacs(accumDsd, accumDsd, srcZ6, coefficients[6]);
const srcZ7 = @increment_dsd_offset(srcZ0, 7, f32);
@fmacs(accumDsd, accumDsd, srcZ7, coefficients[7]);
const srcZ8 = @increment_dsd_offset(srcZ0, 8, f32);
@fmacs(accumDsd, accumDsd, srcZ8, coefficients[8]);
\end{lstlisting}
\caption{Equivalent CSL code. Each \lstinline{@fmacs} instruction takes an output argument and three input arguments. \lstinline{@get_dsd} returns a descriptor, corresponding to a \textit{view} of an array. \lstinline{@increment_dsd_offset} allows to offset the array pointed by an existing descriptor.}
\end{subfigure}

\caption{Applying the stencil along the $Z$ dimension.
\label{lst.z_update}}
\end{figure*}

\subsection{Time integration}

Once the Laplacian has been computed, the time iteration step given in Equation~\ref{eq:minimod-semidisc} can happen. The wavefield from the previous time step is added to the \textit{accumulator} buffer. In reality, this is also done during the stencil computation: as mentioned earlier, a PE receives its own data. Doing so allows to use cycles which would have otherwise been wasted.

The next step is to update the wavefield (per Equation~\ref{eq:minimod-semidisc}), by subtracting the wavefield to the \textit{accumulator} buffer (step 3 in Figure~\ref{fig.xy_update}).

Next, a stimulus, called \textit{source}, needs to be added to a particular cell (with coordinates \textit{(srcX, srcY, srcZ)}) at each time step.
The source value at the current time step is added to the wavefield at offset \textit{srcZ} on the PE with coordinates \textit{(srcX, srcY)}.


%% file: figures/stencil.pdf_tex
\begingroup%
  \makeatletter%
  \providecommand\color[2][]{%
    \errmessage{(Inkscape) Color is used for the text in Inkscape, but the package 'color.sty' is not loaded}%
    \renewcommand\color[2][]{}%
  }%
  \providecommand\transparent[1]{%
    \errmessage{(Inkscape) Transparency is used (non-zero) for the text in Inkscape, but the package 'transparent.sty' is not loaded}%
    \renewcommand\transparent[1]{}%
  }%
  \providecommand\rotatebox[2]{#2}%
  \newcommand*\fsize{\dimexpr\f@size pt\relax}%
  \newcommand*\lineheight[1]{\fontsize{\fsize}{#1\fsize}\selectfont}%
  \ifx\svgwidth\undefined%
    \setlength{\unitlength}{435.42144535bp}%
    \ifx\svgscale\undefined%
      \relax%
    \else%
      \setlength{\unitlength}{\unitlength * \real{\svgscale}}%
    \fi%
  \else%
    \setlength{\unitlength}{\svgwidth}%
  \fi%
  \global\let\svgwidth\undefined%
  \global\let\svgscale\undefined%
  \makeatother%
  \begin{picture}(1,0.94442558)%
    \lineheight{1}%
    \setlength\tabcolsep{0pt}%
    \put(0,0){\includegraphics[width=\unitlength,page=1]{stencil.pdf}}%
    \put(0.86987329,0.38022375){\color[rgb]{0,0,0}\makebox(0,0)[lt]{\lineheight{1.25}\smash{\begin{tabular}[t]{l}$X$\end{tabular}}}}%
    \put(0.12580616,0.10658406){\color[rgb]{0,0,0}\makebox(0,0)[lt]{\lineheight{1.25}\smash{\begin{tabular}[t]{l}$Y$\end{tabular}}}}%
    \put(0.49143529,0.89140599){\color[rgb]{0,0,0}\makebox(0,0)[lt]{\lineheight{1.25}\smash{\begin{tabular}[t]{l}$Z$\end{tabular}}}}%
    \put(0,0){\includegraphics[width=\unitlength,page=2]{stencil.pdf}}%
  \end{picture}%
\endgroup%

%% file: figures/domain.pdf_tex
\begingroup%
  \makeatletter%
  \providecommand\color[2][]{%
    \errmessage{(Inkscape) Color is used for the text in Inkscape, but the package 'color.sty' is not loaded}%
    \renewcommand\color[2][]{}%
  }%
  \providecommand\transparent[1]{%
    \errmessage{(Inkscape) Transparency is used (non-zero) for the text in Inkscape, but the package 'transparent.sty' is not loaded}%
    \renewcommand\transparent[1]{}%
  }%
  \providecommand\rotatebox[2]{#2}%
  \newcommand*\fsize{\dimexpr\f@size pt\relax}%
  \newcommand*\lineheight[1]{\fontsize{\fsize}{#1\fsize}\selectfont}%
  \ifx\svgwidth\undefined%
    \setlength{\unitlength}{641.70389971bp}%
    \ifx\svgscale\undefined%
      \relax%
    \else%
      \setlength{\unitlength}{\unitlength * \real{\svgscale}}%
    \fi%
  \else%
    \setlength{\unitlength}{\svgwidth}%
  \fi%
  \global\let\svgwidth\undefined%
  \global\let\svgscale\undefined%
  \makeatother%
  \begin{picture}(1,0.90730634)%
    \lineheight{1}%
    \setlength\tabcolsep{0pt}%
    \put(0,0){\includegraphics[width=\unitlength,page=1]{domain.pdf}}%
    \put(0.88960703,0.44132876){\color[rgb]{0,0,0}\makebox(0,0)[lt]{\lineheight{1.25}\smash{\begin{tabular}[t]{l}$nz$\end{tabular}}}}%
    \put(0,0){\includegraphics[width=\unitlength,page=2]{domain.pdf}}%
    \put(0.11324642,0.17946149){\color[rgb]{0,0,0}\makebox(0,0)[rt]{\lineheight{1.25}\smash{\begin{tabular}[t]{r}$ny$\end{tabular}}}}%
    \put(0,0){\includegraphics[width=\unitlength,page=3]{domain.pdf}}%
    \put(0.55430378,0.00824069){\color[rgb]{0,0,0}\makebox(0,0)[lt]{\lineheight{1.25}\smash{\begin{tabular}[t]{l}$nx$\end{tabular}}}}%
    \put(0,0){\includegraphics[width=\unitlength,page=4]{domain.pdf}}%
    \put(0.79109506,0.56380889){\color[rgb]{0,0,0}\rotatebox{-90}{\makebox(0,0)[lt]{\lineheight{1.25}\smash{\begin{tabular}[t]{l}MEMORY\end{tabular}}}}}%
    \put(0.49222754,0.11407478){\color[rgb]{0,0,0}\makebox(0,0)[lt]{\lineheight{1.25}\smash{\begin{tabular}[t]{l}FABRIC\end{tabular}}}}%
    \put(0.16506614,0.27564977){\color[rgb]{0,0,0}\rotatebox{-45}{\makebox(0,0)[lt]{\lineheight{1.25}\smash{\begin{tabular}[t]{l}FABRIC\end{tabular}}}}}%
  \end{picture}%
\endgroup%

%% file: figures/z_slice.pdf_tex
\begingroup%
  \makeatletter%
  \providecommand\color[2][]{%
    \errmessage{(Inkscape) Color is used for the text in Inkscape, but the package 'color.sty' is not loaded}%
    \renewcommand\color[2][]{}%
  }%
  \providecommand\transparent[1]{%
    \errmessage{(Inkscape) Transparency is used (non-zero) for the text in Inkscape, but the package 'transparent.sty' is not loaded}%
    \renewcommand\transparent[1]{}%
  }%
  \providecommand\rotatebox[2]{#2}%
  \newcommand*\fsize{\dimexpr\f@size pt\relax}%
  \newcommand*\lineheight[1]{\fontsize{\fsize}{#1\fsize}\selectfont}%
  \ifx\svgwidth\undefined%
    \setlength{\unitlength}{272.60646309bp}%
    \ifx\svgscale\undefined%
      \relax%
    \else%
      \setlength{\unitlength}{\unitlength * \real{\svgscale}}%
    \fi%
  \else%
    \setlength{\unitlength}{\svgwidth}%
  \fi%
  \global\let\svgwidth\undefined%
  \global\let\svgscale\undefined%
  \makeatother%
  \begin{picture}(1,2.04719571)%
    \lineheight{1}%
    \setlength\tabcolsep{0pt}%
    \put(0,0){\includegraphics[width=\unitlength,page=1]{z_slice.pdf}}%
    \put(0.36466278,1.77969924){\color[rgb]{0,0,0}\makebox(0,0)[lt]{\lineheight{1.25}\smash{\begin{tabular}[t]{l}4\end{tabular}}}}%
    \put(0.36466278,0.13461175){\color[rgb]{0,0,0}\makebox(0,0)[lt]{\lineheight{1.25}\smash{\begin{tabular}[t]{l}4\end{tabular}}}}%
    \put(0.74013962,0.98734797){\color[rgb]{0,0,0}\makebox(0,0)[lt]{\lineheight{1.25}\smash{\begin{tabular}[t]{l}$nz$\end{tabular}}}}%
    \put(0.37697879,0.57059666){\color[rgb]{0,0,0}\makebox(0,0)[lt]{\lineheight{1.25}\smash{\begin{tabular}[t]{l}$b$\end{tabular}}}}%
    \put(0,0){\includegraphics[width=\unitlength,page=2]{z_slice.pdf}}%
  \end{picture}%
\endgroup%

%% file: figures/z_update.pdf_tex
\begingroup%
  \makeatletter%
  \providecommand\color[2][]{%
    \errmessage{(Inkscape) Color is used for the text in Inkscape, but the package 'color.sty' is not loaded}%
    \renewcommand\color[2][]{}%
  }%
  \providecommand\transparent[1]{%
    \errmessage{(Inkscape) Transparency is used (non-zero) for the text in Inkscape, but the package 'transparent.sty' is not loaded}%
    \renewcommand\transparent[1]{}%
  }%
  \providecommand\rotatebox[2]{#2}%
  \newcommand*\fsize{\dimexpr\f@size pt\relax}%
  \newcommand*\lineheight[1]{\fontsize{\fsize}{#1\fsize}\selectfont}%
  \ifx\svgwidth\undefined%
    \setlength{\unitlength}{780.59902882bp}%
    \ifx\svgscale\undefined%
      \relax%
    \else%
      \setlength{\unitlength}{\unitlength * \real{\svgscale}}%
    \fi%
  \else%
    \setlength{\unitlength}{\svgwidth}%
  \fi%
  \global\let\svgwidth\undefined%
  \global\let\svgscale\undefined%
  \makeatother%
  \begin{picture}(1,0.58934879)%
    \lineheight{1}%
    \setlength\tabcolsep{0pt}%
    \put(0,0){\includegraphics[width=\unitlength,page=1]{z_update.pdf}}%
    \put(0.15093261,0.00773141){\color[rgb]{0,0,0}\makebox(0,0)[t]{\lineheight{1.25}\smash{\begin{tabular}[t]{c}(a)\end{tabular}}}}%
    \put(0.38152473,0.00773141){\color[rgb]{0,0,0}\makebox(0,0)[t]{\lineheight{1.25}\smash{\begin{tabular}[t]{c}(b)\end{tabular}}}}%
    \put(0.63133356,0.00773141){\color[rgb]{0,0,0}\makebox(0,0)[t]{\lineheight{1.25}\smash{\begin{tabular}[t]{c}(c)\end{tabular}}}}%
    \put(0.89074983,0.00773141){\color[rgb]{0,0,0}\makebox(0,0)[t]{\lineheight{1.25}\smash{\begin{tabular}[t]{c}(d)\end{tabular}}}}%
    \put(0.08351811,0.16991561){\color[rgb]{0,0,0}\makebox(0,0)[rt]{\lineheight{1.25}\smash{\begin{tabular}[t]{r}$z$\end{tabular}}}}%
    \put(0.10818355,0.06615054){\color[rgb]{0,0,0}\makebox(0,0)[rt]{\lineheight{1.25}\smash{\begin{tabular}[t]{r}$z-4$\end{tabular}}}}%
    \put(0.34454049,0.09113102){\color[rgb]{0,0,0}\makebox(0,0)[rt]{\lineheight{1.25}\smash{\begin{tabular}[t]{r}$z-2$\end{tabular}}}}%
    \put(0.59434866,0.11418992){\color[rgb]{0,0,0}\makebox(0,0)[rt]{\lineheight{1.25}\smash{\begin{tabular}[t]{r}$z-3$\end{tabular}}}}%
    \put(0.84992161,0.14493513){\color[rgb]{0,0,0}\makebox(0,0)[rt]{\lineheight{1.25}\smash{\begin{tabular}[t]{r}$z-1$\end{tabular}}}}%
    \put(0,0){\includegraphics[width=\unitlength,page=2]{z_update.pdf}}%
  \end{picture}%
\endgroup%

%% file: experiments.tex
\section{Experimental Evaluation}
\label{sec:eval}

In this section, experimental results of \fd running on a Wafer-Scale Engine are presented.
The scalability and energy efficiency achieved by \fd on this massively parallel platform are discussed.

\subsection{Experimental Configuration}

The experiments are conducted on two platforms: a Cerebras CS-2 equipped with a \wsetwo chip, and a GPU-based platform used as a reference.
The CS-2 is Cerebras' second generation chassis, which uses the 7nm \wsetwo second generation Wafer-Scale Engine. The \wsetwo offers $2.2\times$ more processing elements than the original \wse.
The experiments used a fabric of size $755\times 994$ out of the total 850,000 processing elements of the \wsetwo.
The CS-2 is driven by a Linux server on which no computations take place in the context of this work.
The \wsetwo platform uses Cerebras SDK 0.3.0~\cite{sdk_arxiv}.

The GPU-based platform is \textit{Cypress} from TotalEnergies. It has 4 NVIDIA A100 GPUs, each offering 40 GB of on-device RAM, a 16-core AMD EPYC 7F52 CPU, and 256 GB of main memory. The GPU platform is using CUDA 11.2 and GCC 8.3.1.

Numerical results produced by \fd on \wsetwo are compared to the results produced by Minimod.



\subsection{Weak scaling Experiments}

This section discusses scalability results of \fd on a \wsetwo system.
In order to characterize the scalability of \fd, the grid dimension is modified along the
$X$ and $Y$ dimensions, while the $Z$ dimension (residing in memory) is kept constant to a relevant
value for this type of application. 
The $X$ and $Y$ dimensions are grown up to a size of $755 \times 994$.
Results presented in Table~\ref{tab.weak_scaling} show the throughput achieved on \wsetwo
in Gigacell/s, the wall-clock time required to compute 1,000 time steps on \wsetwo, as well as
timings on a GPGPU provided as baseline. Timing reported in this section correspond to computations taking place on the device only, be it on GPU or \wsetwo.

As can be seen in the table, for all problem sizes,
the wall-clock time required on \wsetwo is constant, meaning that \fd scales nearly perfectly on this platform.
It is crucial to observe that such a reduction in wall-clock time has a significant impact in practice since this type of computations is repeated hundreds of thousands of times in an industrial context.
\fd reaches a throughput of 9872.78 Gcell/s on the largest problem size, which is rarely seen at single system level. This type of throughput is difficult to achieve without using a large number of nodes on distributed-memory supercomputers due to limited strong scalability.

\begin{figure}[h!]
\centering
\includegraphics[scale=.31]{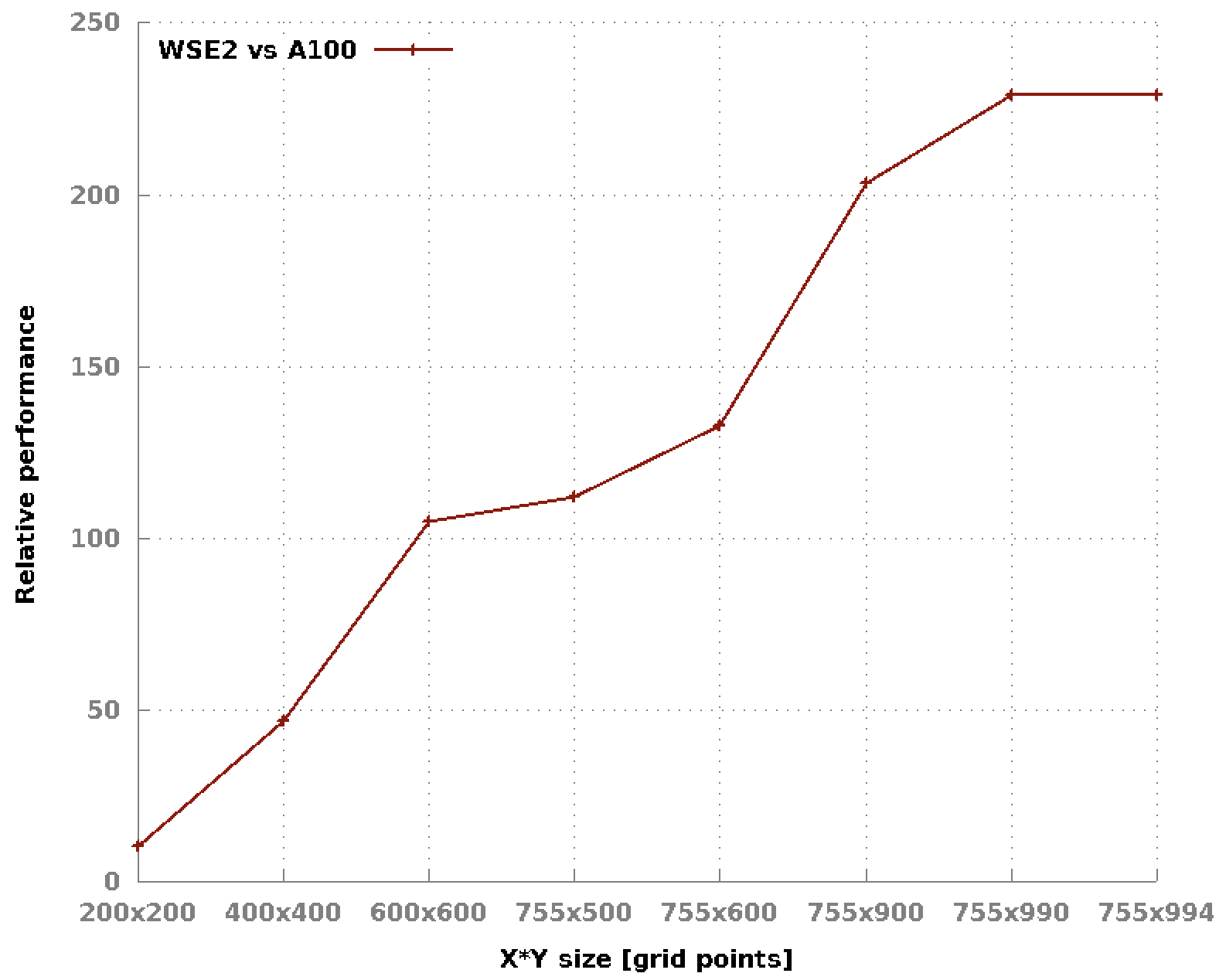}
\caption{Comparisons between implementation on WSE-2 and A100 using elapsed time describe in Table~\ref{tab.weak_scaling}. Fixed $nz = 1000$.
\label{fig.comparison}}
\vspace{-0.15cm}
\end{figure}

Figure~\ref{fig.comparison} depicts the ratio between the elapsed time achieved by the A100-tuned kernel compared to \fd on \wsetwo. As can be seen, when the largest problem is solved (grid size of $755 \times 994 \times 1000$), a speedup of 228x is achieved. While this number shows great potential, it is understood that using multiple GPUs will likely narrow the gap. However, it is unlikely that such a performance gap can be closed entirely, given the strong scalability issues encountered by this kind of algorithm when using a large number of multi-GPU nodes in HPC clusters (\cite{Shimokawabe2017, Anjum2022}).

\begin{figure}[h!]
\centering
\includegraphics[scale=.32]{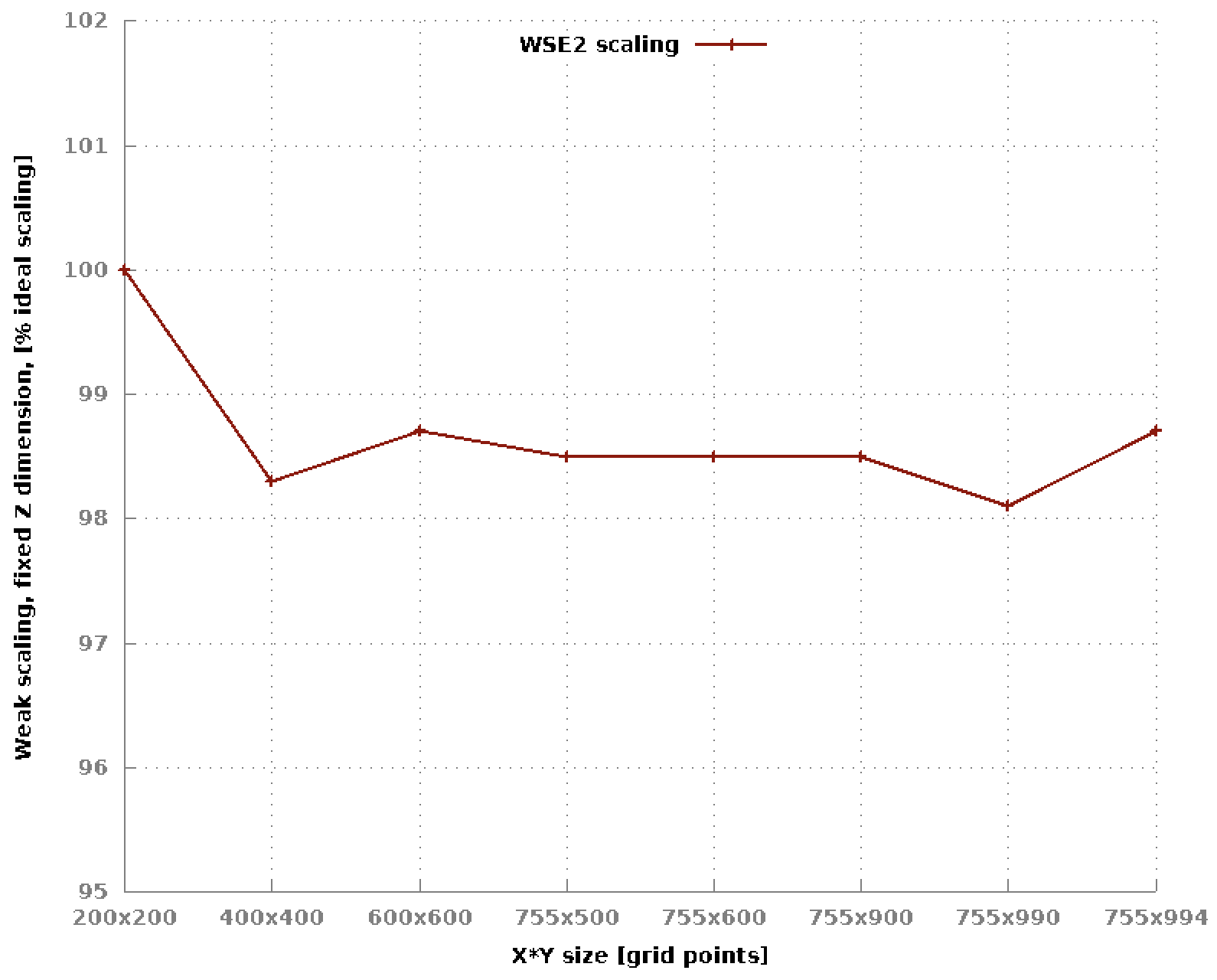}
\caption{Weak scaling under assumption that PE memory is fully utilized ($nz$ = 1000).}
\label{fig.weak}
\end{figure}

\fd shows close to perfect weak scaling on \wsetwo. No matter what the grid size is, the run time stays fairly stable. Taking the 200x200 case as a reference, percentages of the ideal weak scaling for various grid sizes are depicted in Figure~\ref{fig.weak}. As can be seen in the plot, \fd systematically reaches over 98\% of
weak scaling efficiency.
This demonstrates how extremely low latency interconnect coupled with local fast memories can be efficiently leveraged by stencil applications relying on a localized communication pattern.

In the next experiment, the sizes of the $X$ and $Y$ dimensions of the grid are fixed to $nx=755$ and $ny=994$ while the size of the $Z$ dimension $nz$ is varied from $100$ to $1,000$. Results presented in Table~\ref{tab.weak_scaling_mem} show that the throughput increases slightly with $nz$. This indicates that the implementation gains in efficiency due to larger block sizes $b$ and therefore lower relative overheads. More importantly, it confirms that memory accesses do not limit the performance of the implementation, confirming that it is compute-bound.


\begin{table}[ht]
\centering
    \begin{tabular}{ccc|cc|c} \\\toprule
   $nx$ & $ny$ & $nz$ & Throughput & \wsetwo &  A100\\
   & & & Gcell/s & time [s] & time [s] \\\midrule
   200 & 200 & 1000 & 533.64 &  0.0750 & 0.7892 \\
   400 & 400 & 1000 & 2097.60 & 0.0763 & 3.5828 \\
   600 & 600 & 1000 & 4731.53 & 0.0761 & 8.0000 
        \\ \midrule
    755 & 500 & 1000 & 4956.17 & 0.0762 & 8.5499 \\
    755 & 600 & 1000 & 5945.40 & 0.0762 & 10.1362 \\
    755 & 900 & 1000 & 8922.08 & 0.0762 & 15.5070\\
    755 & 990 & 1000 & 9782.14 & 0.0764 & 17.4991\\
    755 & 994 & 1000 & 9862.78 & 0.0761 & 17.4186
    \\ \bottomrule
    \end{tabular}
    \caption{Experimental results for 1,000 time steps for various grid sizes with fixed $nz$.} \label{tab.weak_scaling}
\end{table}

\begin{table}[ht]
\centering
    \begin{tabular}{ccccc} \\\toprule
    $nz$ & $b$  &  Throughput     &   \wsetwo        &   Scaling  \\
     & &    Gcell/s             & time [s]       &       \\\midrule
100  & 100 & 8688.76 &0.8637 &1.0000    \\
200  & 200 & 9303.26 &1.6133 &1.8679    \\
300  & 300 & 9492.89 &2.3716 &2.7458    \\
400  & 400 & 9614.15 &3.1223 &3.6151    \\
500  & 250 & 9786.51 &3.8342 &4.4392    \\
700  & 350 & 9885.04 &5.3143 &6.1531    \\
1000 & 334 & 9936.79 &7.5524 &8.7442    \\
\bottomrule
    \end{tabular}
    \caption{Experimental results, fixed $nx \times ny$ grid dimensions of $755 \times 994$. 100,000 time steps.} \label{tab.weak_scaling_mem}
\end{table}


\subsection{Profiling data}

In the following, various profiling results are provided for \fd, with the objective to provide general insights on \wsetwo-based computations.

Using Cerebras' hardware profiling tool on a $600 \times 600 \times 1000$ grid, the execution of \fd results in an average of 69.6\% busy cycles. The 4 PEs at the corners of the grid have 0 busy cycles since they are not doing any computation. There is an average of 11.6\% idle cycles caused by memory accesses. As expected, the load is extremely balanced, with a standard deviation of 0.8\%. This shows that the hardware is kept busy during the experiment, further confirming the efficiency of the approach proposed in this work.

\begin{figure}[h!]
\centering
\includegraphics[scale=.33]{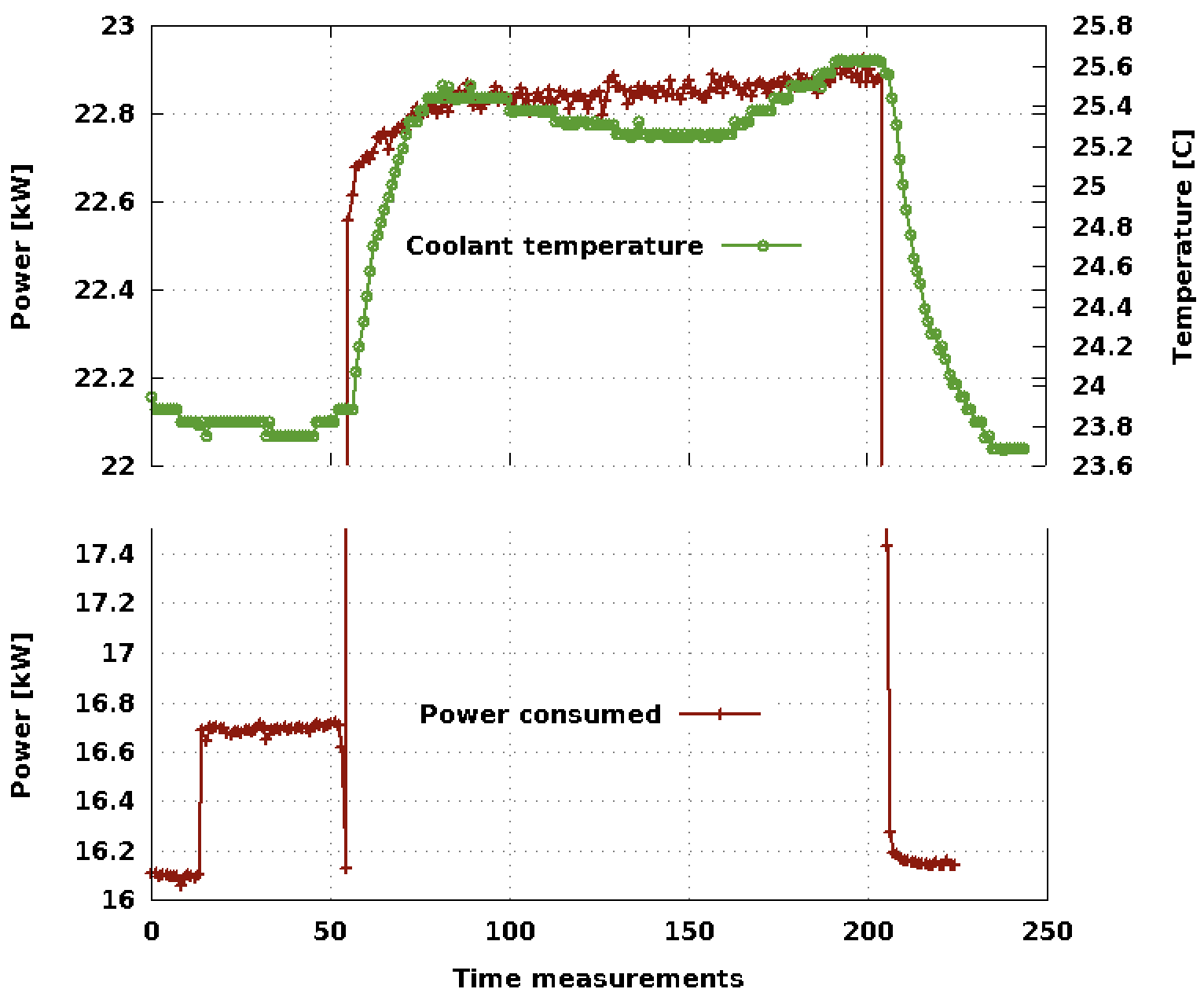}
\caption{Power consumption during experiment with grid $755\times 994\times 1000$ for 10,000,000 time steps, which amounts to 754 seconds. In the plot, time is subsampled by 3.5 seconds. The power baseline is 16.1 kW and peak is 22.9 kW. 
Coolant temperature is also reported, with a baseline of $23.6^{\circ}$C and a peak of $25.6^{\circ}$C.
\label{fig.power}}
\vspace{-0.15cm}
\end{figure}

The power consumption of the CS-2 during a \fd run on a $755 \times 994 \times 1000$ grid is reported in Figure~\ref{fig.power}. In order to record a sufficient number of samples, the run time is extended by setting the number of time steps to 10,000,000, leading to a total run time of 754 seconds. The average power consumption during the execution is 22.8 kW, which corresponds to 
22 GFLOP/W. Such an energy efficiency is hard to find in the literature for a stencil of this order.
 In addition to power consumption, Figure~\ref{fig.power} also depicts the coolant temperature of the CS-2, which uses a closed-loop water-cooling system. During the execution, the coolant temperature rises very moderately from $23.6^{\circ}$C to a peak of $25.6^{\circ}$C.

Altogether, experiments show that the \fd algorithm presented in this study is able to exploit low-latency distributed memory architectures such as the \wsetwo with very high hardware utilization. The application has near-perfect weak scalability and provides significant speedups over a reference GPGPU implementation .

%% file: roofline.tex
\newsavebox{\gpuroofline}

\savebox{\gpuroofline}{%
\includegraphics[scale=0.5]{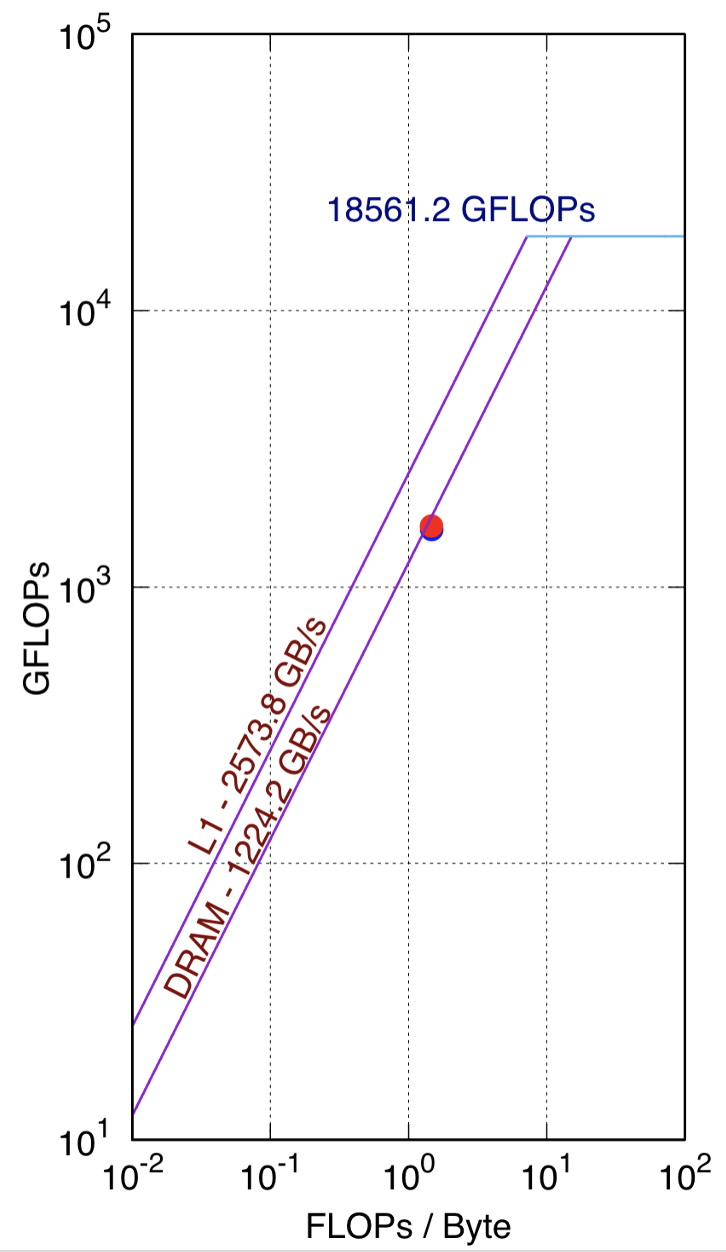}
}

\begin{figure}
\centering
\includegraphics[scale=0.48]{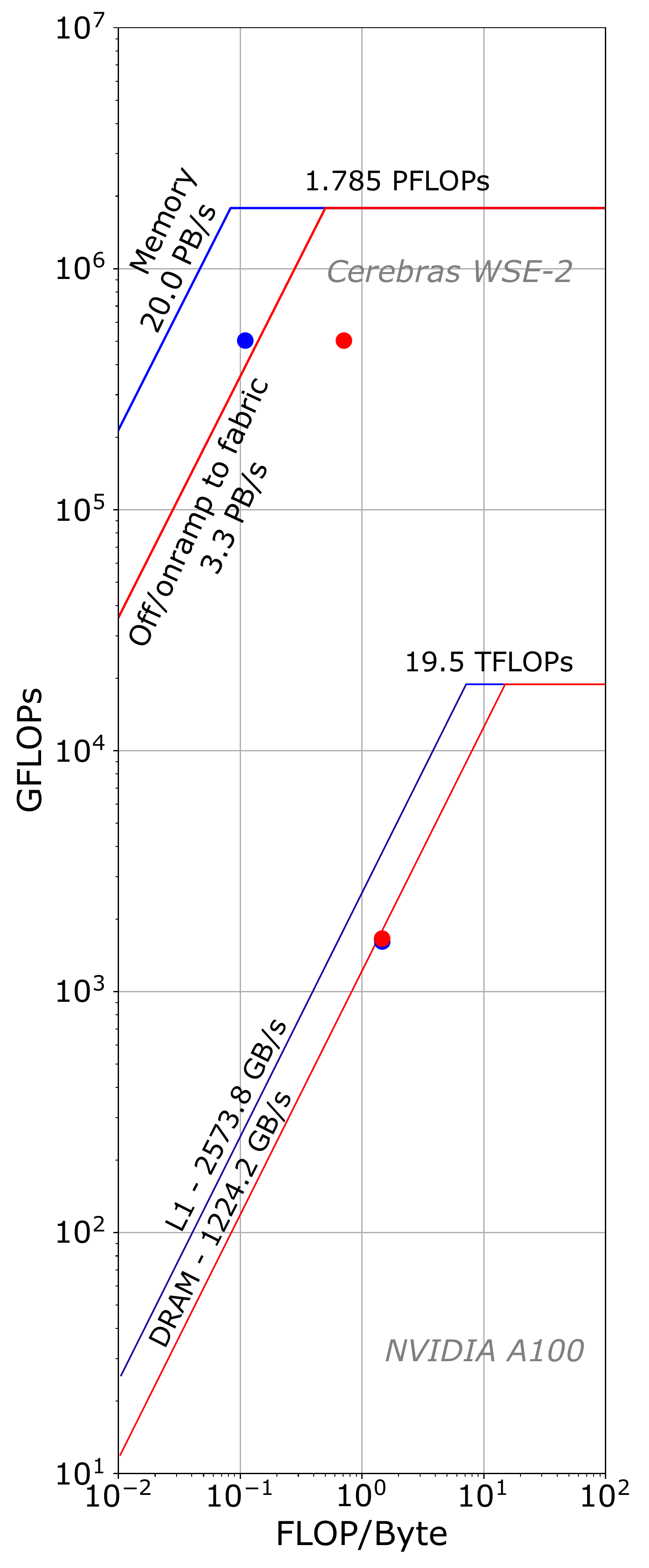}
\caption{Roofline models for \wsetwo and GPGPU-based implementations (in log-log scale) for a $755 \times 994 \times 1000$ grid. Dots represent \fd implementations. 
\\
\wsetwo (top) has two distinct resources: memory and fabric: the leftmost blue dot corresponds to memory accesses, while the red dot to the right corresponds to fabric accesses. Rooflines are given using the same colors. The kernel is clearly in the compute-bound zone for both memory and fabric accesses.
\\
On GPGPU (bottom), red dots and lines correspond to DRAM accesses. L1 cache accesses are depicted in blue. The kernel is clearly in the bandwidth-bound zone. \label{fig.roofline}}
\end{figure}

\section{Roofline model}
\label{sec:roof}

A roofline model~\cite{roofline} is a synthetic view of how many floating point instructions per cycles can be done. This is the \textit{peak} compute capacity of the platform. However, no computation can be done without loading data: the number of 32 bit words that can be accessed per cycle will also impact the peak compute rate. In the case of a \textit{bandwidth-bound} application, the memory will actually determine the performance of the overall
application. In a similar fashion, accesses to the interconnect can be taken into account when the architecture is a distributed memory 
platform.
This ratio between floating point operations (i.e. the ``useful'' 
operations) and the volume of data coming from/going to a given resource, for instance memory,  is referred to as \textit{arithmetic intensity} (measured in FLOP/byte).


The \wsetwo is a SIMD-centric multiprocessor providing up to 4 simultaneous 16 bit floating point instructions per cycle.
In the context of this work, only 32 bit floating point operations are used.
The peak number of floating point operations per second (FLOPs) is represented by the horizontal line at the top of Figure~\ref{fig.roofline}.

The \wsetwo does not have a complex cache memory hierarchy: each PE has a single local memory accessed directly. Four 32 bit packets can be accessed from memory per cycle, and up to two packets can be stored to memory per cycle. 
In memory intensive application, memory limits the peak 
achievable performance. This corresponds to the slanted blue line at the top of Figure~\ref{fig.roofline}. 

Each PE is connected to its router by a bi-directional link able to move 32 bits per cycle in each direction (referred to as ``off/onramp bandwidth'' in the plot). The router is connected to other routers by 4 bi-directional links, each moving 32 bits per cycle. This corresponds to the slanted red line at the top of Figure~\ref{fig.roofline}.  



%
For each cell, the stencil computation in \fd involves 25 multiplies and 25 adds. In
addition to that, \fd requires a subtraction between the previous time step and the current time step. This corresponds to a total of 51 floating point instructions per cell.
As explained in Section~\ref{sec:wse}, due to architecture constraints, only the computation of the stencil along the $Z$ dimension can be done using FMAs. For the $X$ and $Y$ dimensions, the implementation relies on separate FMUL and FADD instructions. The final value of the current time step is computed using a FSUB operation.
A summary of the instructions used in \fd, floating point operations per cell, memory traffic, fabric traffic, and instruction count per cell is given in Table~\ref{tab:fd_instructions}.

\begin{table}
    \centering
    \begin{adjustbox}{width=\linewidth}
    \begin{tabular}{c||c|c|c}
        \hline
        \textbf{Operation} & FLOP & Mem. traffic & Fabric traffic\\
        \hline
        17 FMUL & 1 & $1/b$ load, 1 store  & 1 load\\ 
        17 FADD  & 1 & 2 loads, 1 store  & 0\\ 
        8 FMA   & 2 & 2 + $1/(b)$ loads, 1 store  & 0\\ 
        1 FSUB  & 1 & 2 loads, 1 store  & 0\\ 
        \hline
    \end{tabular}
    \end{adjustbox}
    \caption{Instruction and memory access counts of the \fd implementation on \wsetwo}
    \label{tab:fd_instructions}
    \vspace{-0.15cm}
\end{table}

On \wsetwo, \fd computes 57 floating point operations per cell, 51 of which are required by the algorithm. Extra operations are due to hardware constraints. The number of operations strictly required by the algorithm is used in all performance numbers. These 51 floating point operations require 112 load and store of 32 bit words from/to memory, and 17 loads from fabric. This leads to an \textit{arithmetic intensity} of 0.11 with respect to memory accesses, and 0.75 with respect to fabric transfers.

On \wsetwo, a $755 \times  994 \times 1000$ grid is computed in 0.0761s (see Table~\ref{tab.weak_scaling}). This leads to a flop rate of 670.3
MFLOPs per PE, and an aggregated performance of 503 TFLOPs for the entire grid of PEs used by this problem size.
The roofline model of the \wsetwo, depicted in Figure~\ref{fig.roofline}(top), indicates that \fd is \textit{compute bound} thanks
to the extremely fast local memory. This is quite remarkable, and confirms the weak scaling results given in Section~\ref{sec:eval}. The application is \textit{communication/memory bound}
on most architecture, such as the GPU platform used in this study (roofline model depicted in Figure~\ref{fig.roofline}(bottom)). Note that different optimizations lead to different arithmetic intensities.